\documentclass[5p,sort&compress]{elsarticle}

\usepackage{graphicx}

\usepackage[hidelinks]{hyperref}

\usepackage{comment}
\usepackage[squaren]{SIunits}
\usepackage{bm,amsmath,amsfonts,braket,todonotes,multirow,dcolumn}
\usepackage{ wasysym }

\usepackage{graphicx}
\usepackage{caption}

\usepackage{subcaption}
\usepackage{cleveref}

\begin{document}

\title{EMCD with an electron vortex filter: Limitations and possibilities}

\author[ifp,ustem]{T. Schachinger \corref{cor1}}
\ead{thomas.schachinger@tuwien.ac.at}
\author[ustem,mcmaster]{S. L\"offler}
\author[ustem]{A. Steiger-Thirsfeld}
\author[ifp,ustem]{M. St\"oger-Pollach}
\author[ifw]{S. Schneider}
\author[ifw]{D. Pohl}
\author[ifw]{B. Rellinghaus}
\author[ifp,ustem]{P. Schattschneider}
\cortext[cor1]{Corresponding author}

\address[ifp]{Institute of Solid State Physics, TU Wien, Wiedner~Hauptstra{\ss}e~8-10, 1040 Vienna, Austria}
\address[ustem]{University Service Centre for Transmission Electron Microscopy, TU Wien, Wiedner Hauptstra{\ss}e 8-10, 1040 Wien, Austria}
\address[mcmaster]{Department of Materials Science and Engineering, McMaster University,1280 Main Street West, Hamilton Ontario  L8S 4L8, Canada}
\address[ifw]{Institute for Metallic Materials, IFW Dresden,  P.O. Box 270116, D-01171 Dresden, Germany}

\begin{abstract}
We discuss the feasibility of detecting spin polarized electronic transitions with a vortex filter. This approach does not rely on the principal condition of the standard energy loss magnetic chiral dichroism (EMCD) technique, the precise alignment of the crystal, and thus paves the way for the application of EMCD to new classes of materials and problems. The dichroic signal strength in the L$_{2,3}$-edge of ferromagnetic cobalt is estimated on theoretical grounds. It is shown that magnetic dichroism can, in principle, be detected. However, as an experimental test shows, count rates are currently too low under standard conditions.

\end{abstract}

\maketitle

\section{Introduction}

The discovery in 2006 that magnetic chiral dichroism can be observed in the transmission electron microscope (TEM)~\cite{SchattNature2006} provided an unexpected alternative to X-ray circular dichroism (XMCD) in the synchrotron. Energy-loss magnetic chiral dichroism (EMCD)\footnote{Sometimes also called {\it electron} magnetic chiral dichroism. Properly speaking, this is incorrect because it is not the electron but the energy loss signal that shows dichroism.} has seen tremendous progress since then~\cite{VerbeeckUM2008, WarotJAP2010, WangZhongYuEtAl2013}, achieving  nanometre resolution~\cite{SchattschneiderPRB2008}, and even sub-lattice resolution~\cite{SchattPRB2012,EnnenLoefflerKuebelEtAl2012}. 

The discovery of electron vortex beams (EVBs)~\cite{UchidaNature2010,VerbeeckNature2010} has spurred efforts to use them for EMCD because of their intrinsic chirality. 
In spite of much progress in the production and application of vortex beams~\cite{McMorranScience2011, BliokhSchattschneiderVerbeeckEtAl2012,SchattPRB2012,LoefflerSchattschneider2012,VerbeeckTianVanTendeloo2013,GuzzinatiSchattschneiderBliokhEtAl2013,LubkClarkGuzzinatiEtAl2013}, it soon became clear that atom-sized vortices incident on the specimen are needed for EMCD experiments~\cite{SchattschneiderLoefflerStoeger-PollachEtAl2014,RuszBhowmick2013,RuszBhowmickErikssonEtAl2014}. Attempts to produce such beams and to use them for EMCD measurements did not show an effect so far~\cite{PohlSchneiderRuszEtAl2015}. Nevertheless, faint atomic resolution EMCD signals have been shown without the need for atom-sized EVBs using intelligent shaping of the incident wavefront with a C$_s$ corrector instead~\cite{RuszPRL113,IdroboRuszMcGuireEtAl2015}.

The fact that orbital angular momentum (OAM) can be transfered to the probing electron when it excites electronic transitions to spin polarized final states in the sample manifests itself in a vortical structure of the inelastically scattered probe electron. 
 The latter could be detected by a holographic mask after the specimen. Using a fork mask as chiral filter is already established in optics.~\cite{FicklerLapkiewiczPlickEtAl2012,Bazhenov1990,SteigerBernetRitsch-Marte2013}. This ansatz opens up the possibility to measure magnetic properties of amorphous materials (or multiphase materials including both crystalline and amorphous magnetic phases~\cite{Herzer2013}), since the specimen no longer needs to act as a crystal beam splitter itself. Also crystalline specimens could benefit from using the vortex filter setup and its inherent breaking of the Bragg limitation, when, for example, substrate reflections overlap with the two EMCD measurement positions which would diminish the EMCD signal strength.

\section{Principle}

Dealing with transition metals, dichroism measurements typically involve 2p-core to d-valence excitations at the L$_{2,3}$ ionization edges. The L$_{2,3}$-edges are used, due to their strong spin-orbit interaction in the initial state. Besides their dichroic signal is an order of magnitude higher compared to using K-edges, which were originally used in X-ray magnetic circular dichroism measurements to show the dichroic effect~\cite{WuStoehrHermsmejerEtAl1992,VerbeeckNature2010,SchutzPRL87}.  
The most dominant contribution to the ionisation edges are electric dipole-allowed transitions. Higher multi-pole transitions show low transition amplitudes contributing less than \unit{10}{\%} at scattering angles of $< \unit{20}{\milli\rad}$ relevant in electron energy loss spectrometry (EELS)~\cite{Manson1972,LoefflerEnnenTianEtAl2011,AuerhammerRez1989}. 

In case of an L-edge dipole-allowed transition which changes the magnetic quantum number of an atom by $\mu$, an incident plane wave electron transforms into an outgoing wave~\cite{SchattschneiderPRB2010}
\begin{equation}
\psi_{\mu}({\bf r})=e^{-i\mu\varphi_r} f_\mu(r)
\label{Atom}
\end{equation}
where $\varphi_r$ is the azimuthal angle, and 
\begin{equation}
f_\mu(r)=
\frac{i^{\mu}}{2 \pi}q_E^{1-|\mu|} \int_0^\infty \frac{q^{1+|\mu|} J_{|\mu|}(q r) \langle
j_1(Q) \rangle_{ELSj}}{Q^3} dq,
\label{psi2}
\end{equation}
with $\langle
j_1(Q) \rangle_{ELSj}$ the matrix element of the spherical Bessel function between the initial and final radial atomic wave functions, and $Q=\sqrt{q^2+q_E^2}$. Here, $q$ is the transverse scattering vector that relates to the experimental scattering angle $\theta$ as $q=k_0 \theta$, and $\hbar q_E=\hbar k_0 \theta_E$ is the scalar difference of linear momenta of the probe electron before and after the inelastic interaction, also known as the characteristic momentum transfer in EELS~\cite{RoPiP_v72_i1_p25}. The characteristic scattering angle $\theta_E$ is given by $\theta_E\sim\Delta E/2E_0$, with $\Delta E$ being the threshold energy of the dipole-allowed L-edge and $E_0$ the primary beam energy.
 
\begin{figure}[ht]
	\centering
	\includegraphics[width=90mm]{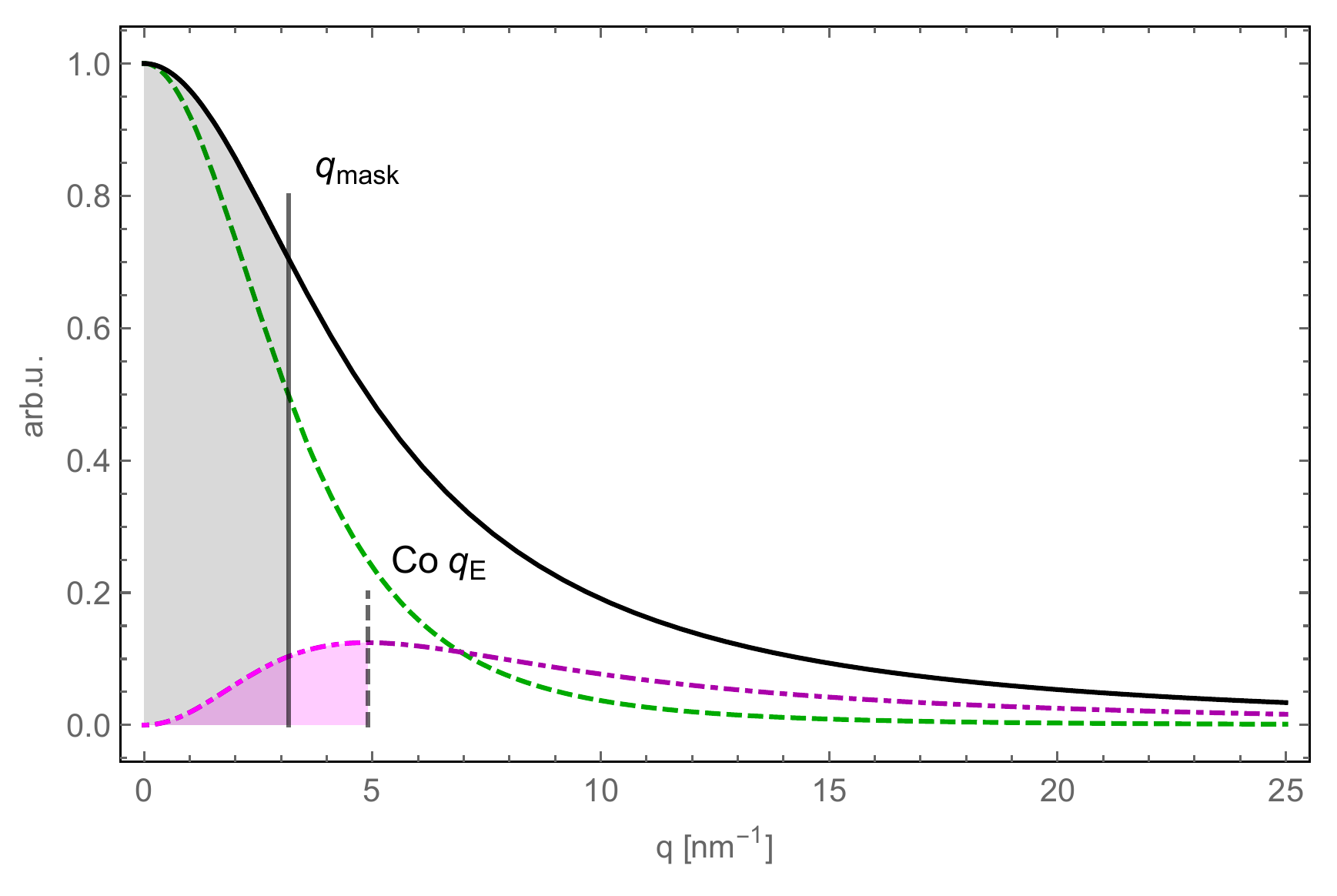} 
	\caption{Scattering profiles of $|\tilde\psi_0|^2$ (green dashed line) and $|\tilde \psi_{\pm 1}|^2 $ (magenta dot dashed line) and their sum  $\sum_{\mu=-1,0,1} |\tilde\psi_\mu|^2$ (black solid line), giving the Lorentz profile for non-magnetic isotropic transitions in momentum space for the Co $L_3$-edge. The radius of the fork mask used in the proof-of-principle experiment $q_{mask}$ is indicated by the grey shaded area terminated by the full vertical line, $q_E$ is indicated by the dashed vertical line and the magenta shaded area.}
	\label{fig:profile}
\end{figure}
The dichroic signal in the diffraction plane (DP) can readily be calculated via Fourier transforming Eq.~\ref{Atom}. According to a  theorem for the Fourier-Bessel transform of a function of azimuthal variation $e^{-i \mu \varphi}$~\cite{AbramowitzStegun1965}, 
one has
\begin{equation}
\tilde \psi_{\mu}({\bf q})= 
\frac{i^{\mu}}{2 \pi} e^{-i\mu\varphi_q}\int_0^{\infty} f_{\mu}(r) J_{|\mu|}(q r)\,  r dr .
\label{FT2}
\end{equation}
The outgoing electron in the DP still carries topological charge $\mu$, showing that the wave function is topologically protected.
The radial intensity profiles $|\tilde \psi_{\mu}({q})|^2$ for the possible transitions with $\mu \in \{-1, 0, 1\}$ and their sum $\sum_{\mu=-1,0,1} |\tilde\psi_\mu|^2$, which represents the Lorentz profile for non-magnetic isotropic transitions, for the Co L$_3$-edge are shown in Fig.~\ref{fig:profile}. Note that in Fig.~\ref{fig:profile}, as well as in all the following simulation results, the parameters used have been adopted to the proof-of-principle experiment given in Sec.~\ref{sec:Experimental}. 
  
\begin{figure*}[]
	\centering
	\includegraphics[width= \textwidth]{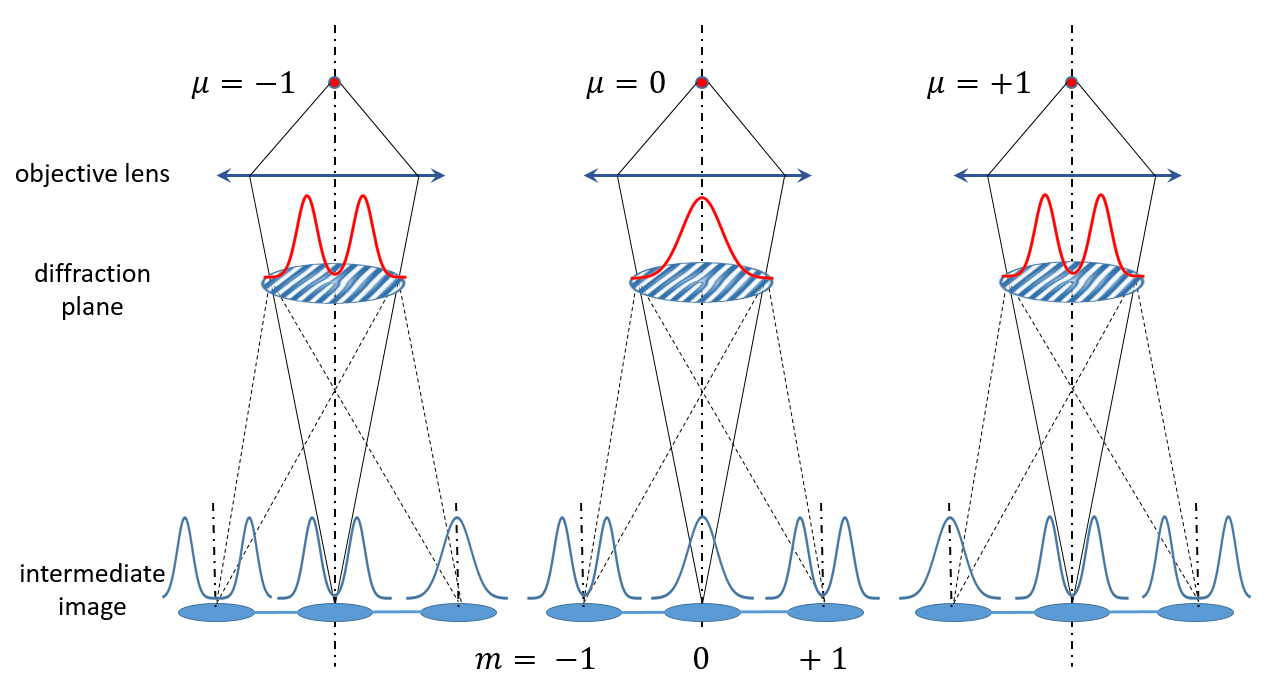} 
	\caption{Principle of the vortex filter (ideal geometry): The red dot represents the scattering centre, i.e. the atom in the object plane, with its three scattering channels, $\mu=\pm1,0$. The resulting vortices are then incident on a fork mask in the far-field, adding topological charges $m \in \mathbb{Z}$, thereby creating a line of vortices of topological charges $\mu+m$ in the image plane. Subsequently, the EMCD signal can be derived from the difference in the vortex orders $m=\pm1$, according to Eq.~\ref{Im}.}
	\label{fig:principle}
\end{figure*}

A fork mask in the DP adds topological charges $m \in \mathbb{Z}$ to the incident beam of topological charge $\mu$. Due to the grating nature of the fork mask, the m-dependent deflections are separated by $2\theta_{Bragg} = \lambda/g$, where $\lambda$ is the electrons' wavelength and $g$ is the fork mask periodicity. Thus, such a mask creates a line of vortices of topological charge  $m+\mu$ in the image plane, see Fig.~\ref{fig:principle}.
The radial profiles in the image plane are given by the back-transform of Eq.~\ref{FT2} with the respective vortex order $m$ added by the mask:
\begin{equation}
 \psi_{m \mu}({\bf r})= 
\frac{i^{m+\mu}}{2 \pi}\, e^{-i(m+\mu)\varphi_r}\int_0^{q_{mask}} \tilde \psi_\mu(q) J_{|m+\mu|}(q\, r)\,  q \,dq 
\label{FT3}
\end{equation}
where $q_{mask}=k_0 \theta_{mask}$ is given by the mask aperture limiting the maximum momentum transfer.
The respective intensities are azimuthally symmetric with distinct radial profiles. Fig.~\ref{fig:principle} shows schematically the central three vortices for the three dipole-allowed transition channels. Note that the central vortex ($m=0, |\mu|=1$) does not show a difference for up and down spin polarization. This is the reason why such transitions cannot be distinguished with standard EELS. Fig.~\ref{fig:principle} only describes the situation where one transition channel is present.

\begin{figure}[ht]
	\centering
	\includegraphics[width= 90mm]{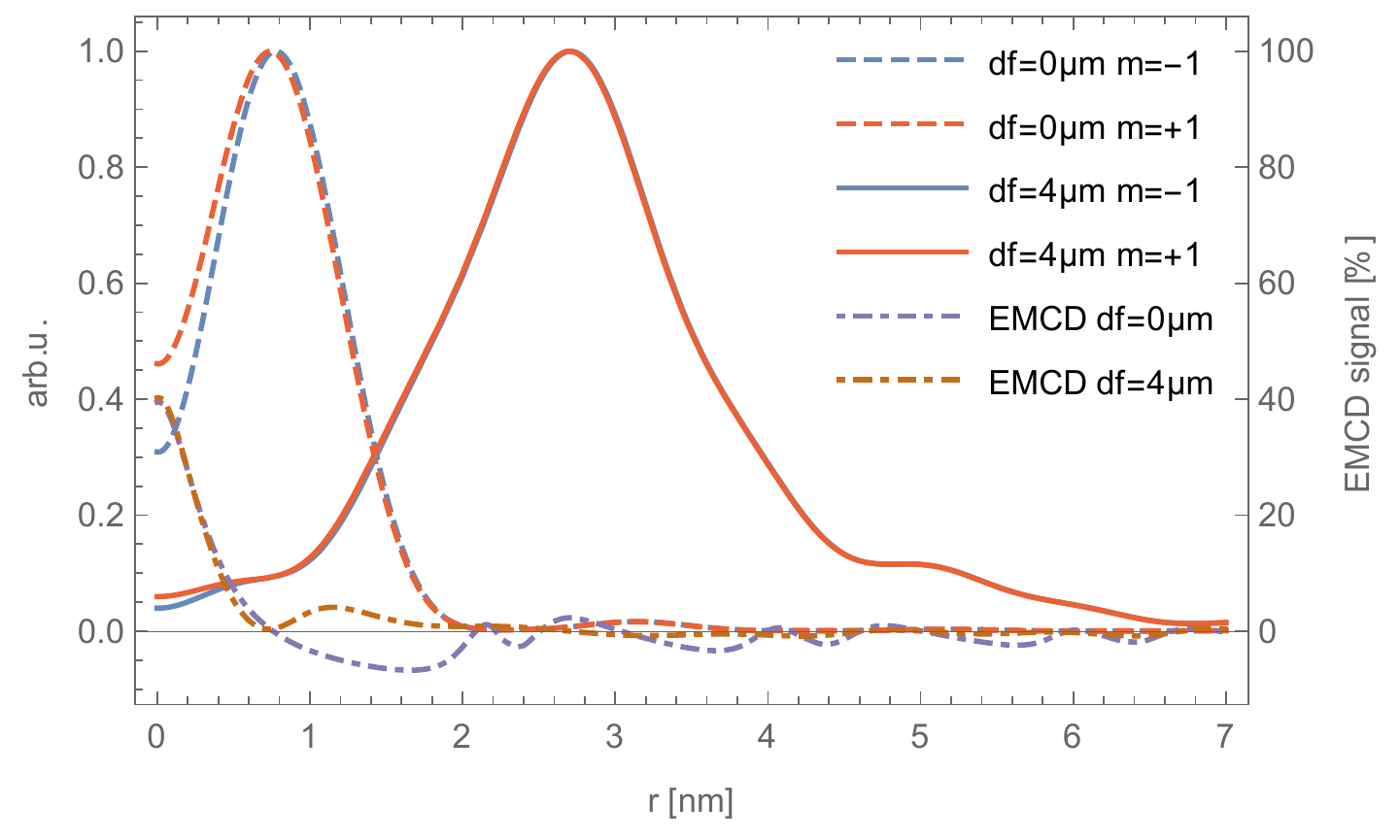} 
	\caption{Radial intensity profiles of a $m=\pm 1$ filtered image of a single atomic ionisation, for the fully spin polarized case according to Eq.~\ref{Im}, with a defocus of \unit{0}{\micro\meter} (dashed lines) and \unit{4}{\micro\meter} (solid lines), respectively. The resulting EMCD signal is given by the dot-dashed curves, according to Eq.~\ref{EMCD}, and amounts to $\sim 40\%$ in the central region of the vortices for both defocus values.}
	\label{fig:CoCoherent}
\end{figure}

For several transition channels at the same energy, the outgoing probe electron is in a mixed state, described by the reduced density matrix~\cite{SchattschneiderNelhiebelJouffrey1999,LoefflerSchattschneider2012a} and the path of rays cannot be visualized in such an easy way. Note that the total intensity is the trace of the matrix, i.e. the sum over all intensities in the respective channels.
For fully spin-polarized systems 
\begin{equation}
I_m=\sum_{\mu=-1}^1 C_m^\mu\,  |\psi_{m \mu}|^2
\label{Im}
\end{equation}
where the $C_m^\mu $ are derived from the Clebsch-Gordan coefficients~\cite{SchattPRB2012,Loffler201339} and given in Tab.~\ref{tab:CGCoefficients}.
\begin{table*}[ht]
\centering
\begin{tabular}{cccccccccc}
\hline
\multicolumn{1}{|c||}{$m$} & \multicolumn{3}{c|}{-1}                                 & \multicolumn{3}{c|}{0}                  & \multicolumn{3}{c|}{+1}                   \\ \hline 
\multicolumn{1}{|c||}{$\mu$} & \multicolumn{1}{c|}{-1} & \multicolumn{1}{c|}{0} & \multicolumn{1}{c|}{+1} & \multicolumn{1}{c|}{-1} & \multicolumn{1}{c|}{0} & \multicolumn{1}{c|}{+1} & \multicolumn{1}{c|}{-1} & \multicolumn{1}{c|}{0} & \multicolumn{1}{c|}{+1}     \\ \hline
\multicolumn{1}{|c||}{$m+\mu$} & \multicolumn{1}{c|}{-2} & \multicolumn{1}{c|}{-1} & \multicolumn{1}{c|}{0} & \multicolumn{1}{c|}{-1} & \multicolumn{1}{c|}{0} & \multicolumn{1}{c|}{+1} & \multicolumn{1}{c|}{0} & \multicolumn{1}{c|}{+1} & \multicolumn{1}{c|}{+2}     \\ \hline
\multicolumn{1}{|c||}{$C_m^\mu$} & \multicolumn{1}{c|}{0.278} & \multicolumn{1}{c|}{0.222} & \multicolumn{1}{c|}{0.167} & \multicolumn{1}{c|}{0.278}  & \multicolumn{1}{c|}{0.222}  & \multicolumn{1}{c|}{0.167}  & \multicolumn{1}{c|}{0.278}  & \multicolumn{1}{c|}{0.222}  & \multicolumn{1}{c|}{0.167}\\ \hline 
\end{tabular}
\caption{Coefficents $C_m^\mu$ for the $L_3$-edge ($j=3/2$) taken from~\cite{SchattPRB2012}. The weighting factors for the transitions when the final states are completely up-spin polarized show an asymmetry for $m+\mu=0$, i.e. in the centre of the $m=\pm1$ vortices. 
}
\label{tab:CGCoefficients}
\end{table*}
In Fig.~\ref{fig:CoCoherent}, the resulting radial intensity profiles are drawn for the coherent case with no inelastic source broadening added (which will be discussed in more detail at the end of this section). The radial extension of the intensity profile in Fig.~\ref{fig:CoCoherent} is considerably broader than it is directly at the scattering centre, due to the limited extent $q_{mask}$ of the vortex filtering mask shown in Fig.~\ref{fig:principle}.  

In this geometry, we define the EMCD signal as the relative difference of the intensities, from vortices with $m=\pm 1$
\begin{equation}
	\text{EMCD}=2 \cdot \frac{I_{+1}-I_{-1}}{I_{+1}+I_{-1}}.
	\label{EMCD}
\end{equation}
The  EMCD signal is a function of the radius which has been omitted for clarity here. It depends on the topological charge of the spin polarized transition $\mu$.
It is the difference of radial profiles for vortex orders $m=-1$ and $m=1$.

In practice, there are two technical problems with the setup shown in Fig.~\ref{fig:principle}. On the one hand, placing the vortex filtering fork mask in the DP is not straightforward, because, due to the limited space in the pole piece gap, strip apertures are used which cannot be loaded with conventional \diameter\unit{3}{\milli\meter} frame apertures. Even though there are proposals to use spiral-phase-plates in the DP, e.g. to determine chiral crystal symmetries and the local OAM content of an electron wave~\cite{JuchtmansVerbeeck2015,JuchtmansVerbeeck2016}, to date no successful implementation of a vortex mask in the DP of a TEM has been shown.
On the other hand, the final image in the selected area aperture (SAA) plane (intermediate image plane) of the atom sized vortices would be too small to be resolvable at all.
Both facts make this direct approach problematic. 
The second obstacle can be removed by defocussing the lens, observing broader vortices.
The first one can be overcome by positioning the fork mask in a relatively easily accessible position, which is the SAA holder. 

\begin{figure}[ht]
	\centering
	\includegraphics[width=90mm]{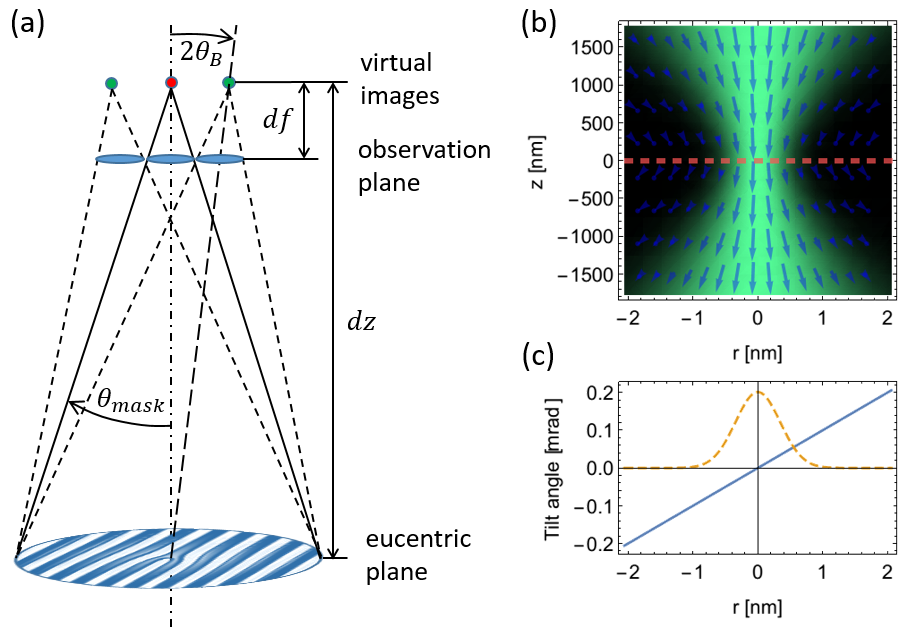} 
	\caption{Scattering geometry, intensity, phase gradient and tilt angles of the wavefront of the incident focused probe: (a) Inserting a fork mask in the SAA creates a virtual image of the fork mask in the eucentric plane. The specimen is lifted by $dz$ such that the fork mask is in the far field, subtending an angle $\theta_{mask}$. It creates a linear series of atom sized vortices. Their virtual images in the object plane, separated by $2\theta_B$, are sketched in green here. The observation plane is $df$ below the object plane, in order to obtain sufficient radial resolution. (b) Intensity of the focused probe of a single electron being incident on the specimen (the red atom in (a)), given in green, and its phase gradient represented by the blue arrows. The red rectangles represent atomic columns, with a spacing of \unit{250}{\pico\meter} and a thickness of \unit{70}{\nano\meter}. Qualitatively it can be seen that the atomic columns practically see a plane wavefront (arrows aligned parallel to the atomic columns) because of the relatively high Rayleigh range $\unit{\sim600}{\nano\meter}$ of the beam.
(c) quantifies the tilt of the incident wavefront. The solid blue line represents the tilt angle of the electron wavefront at \unit{35}{\nano\meter} above the focus, whereas the dashed orange line indicates the lateral beam profile at that position. At the beams' waist the tilt angles are as low as $\unit{70}{\micro\radian}$, justifying the assumption of an incident plane wave.}
	\label{fig:geometry}
\end{figure}

\begin{figure*}[ht]
	\centering
	\includegraphics[width=1. \textwidth]{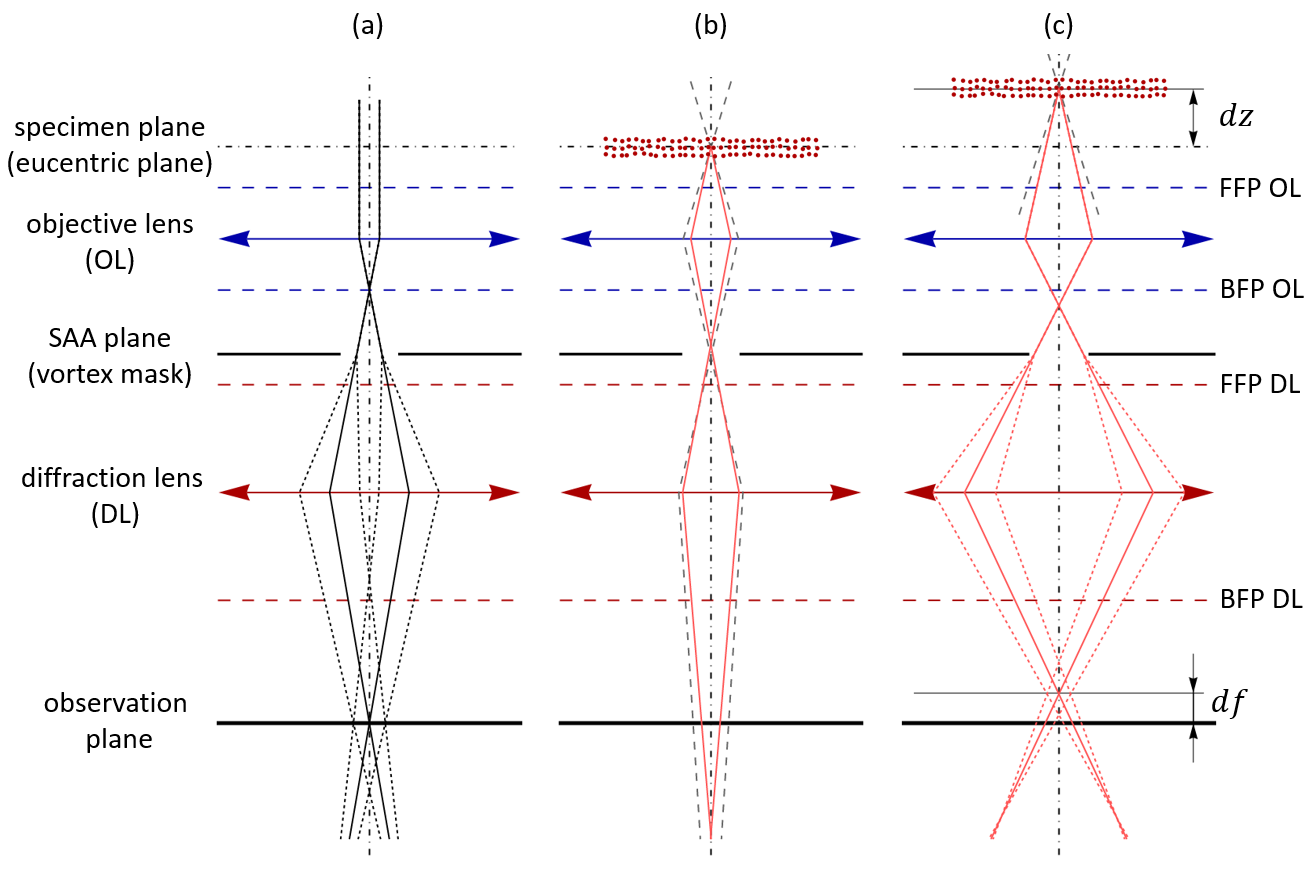} 
	\caption{Ray diagrams (not to scale) of (a) a standard TEM diffraction setup, (b) a standard STEM setup and (c) the EMCD vortex filter setup. Full and dashed black lines represent rays of elastically scattered electrons whereas red lines depict inelastically scattered ones.}
	\label{fig:RayDiagrams}
\end{figure*}

These ideas lead to a scattering geometry that can be easily set up in conventional TEMs, see Fig.~\ref{fig:geometry}a and Fig.~\ref{fig:RayDiagrams}c. Here, the fork mask is positioned in the SAA holder, creating a demagnified virtual image in the eucentric plane with small lattice constant. Additionally, the specimen is lifted in height by $dz$ and the C2 condenser lens is adjusted such that a focused probe is incident on the specimen. 

Note that focusing the beam onto the specimen guarantees that the probability density current is mostly aligned parallel to the optical axis all over the illuminated area such that the scattering "light cones" all point in the same direction towards the vortex filter mask. This is due to the fact that the Rayleigh range of the incident beam\footnote{Note that the Rayleigh range was determined using the diffraction limited spot size of the C2 aperture, the final probe diameter is much larger due to incoherent source broadening.} is of the order of \unit{600}{\nano\meter} (convergence semi-angle \unit{3.8}{\milli\radian}) which is much larger than the sample thickness, in our case $\unit{\sim70}{\nano\meter}$, and thus the incident wavefronts are almost flat everywhere inside the specimen, see Fig.~\ref{fig:geometry}b. How much they actually deviate from that assumption is estimated in Fig.~\ref{fig:geometry}c. It can be seen that the tilt angle at a radial position of \unit{0.7}{\nano\meter} in the entrance plane amounts to $\unit{\sim70}{\micro\radian}$ which can be considered negligible compared to the characteristic scattering angle $\theta_E \sim \unit{1.95}{\milli\radian}$ of Co.
Furthermore only a small fraction of atoms in the illuminated specimen area actually sees such "high" tilt angles of the incident wavefront, most of the illuminated atoms do see a much less tilted wavefront\footnote{Neglecting the crystal field and channelling effects.}.    

Lifting the specimen ensures that the (virtual) fork mask is now in the far field of the excited atom and creates a series of images of the ionization process as depicted in Fig.~\ref{fig:geometry}a. Practically, this setup is comparable to a standard STEM geometry but with the specimen lifted far off the eucentric plane. For better understanding the scattering setup, Fig.~\ref{fig:RayDiagrams} compares the standard TEM setup in diffraction, Fig.~\ref{fig:RayDiagrams}a, and the standard STEM setup, Fig.~\ref{fig:RayDiagrams}b, to the setup proposed here, Fig.~\ref{fig:RayDiagrams}c.
Note that there are slight changes in the focal position of elastic- to inelastically scattered electrons in Fig.~\ref{fig:RayDiagrams}b.
It can be seen that when the vortex filter mask is placed in the SAA holder diffracted beams (small dashing) emerge from the vortex mask in Fig.~\ref{fig:RayDiagrams}a,c but not in Fig.~\ref{fig:RayDiagrams}b because there the image of the inelastically scattered electrons in the SAA plane is much smaller than the grating periodicity such that the vortex mask is not illuminated. Lifting the specimen by $dz$ in Fig.~\ref{fig:RayDiagrams}c ensures that the vortex filter mask is properly illuminated.
Moreover, due to electron optical reasons this lifting is essentially reducing the size of the first image of the focused probe on the (lifted) specimen, comparable to the reduction of the effective source size in the condenser system of a TEM by adjusting the C1 lens excitation.

The dichroic signal is strong in the centre of the vortices but difficult to observe because of their extension of only about 1 nm, as seen in Fig.~\ref{fig:CoCoherent}. Therefore, the observation plane is set at a defocus $df$ (here \unit{4}{\micro\meter}) from the specimen (preferably with the diffraction lens setting) to enhance the visibility of the dichroic signal. This can be understood from Fig.~\ref{fig:geometry}a and Fig.~\ref{fig:RayDiagrams}c. Virtual images (green) of the object intensities are observed at a defocus $df$, making the distribution broader, such that the maximum of the vortices' radial intensity distribution, where no dichroic signal is expected to occur, moves towards higher radii. The orders $m \in \{-1, 0, 1\}$,  with an angular separation of $2\theta_B$ are shown in Fig.~\ref{fig:geometry}a.

\begin{figure}[ht]
	\centering
	\includegraphics[width=90mm]{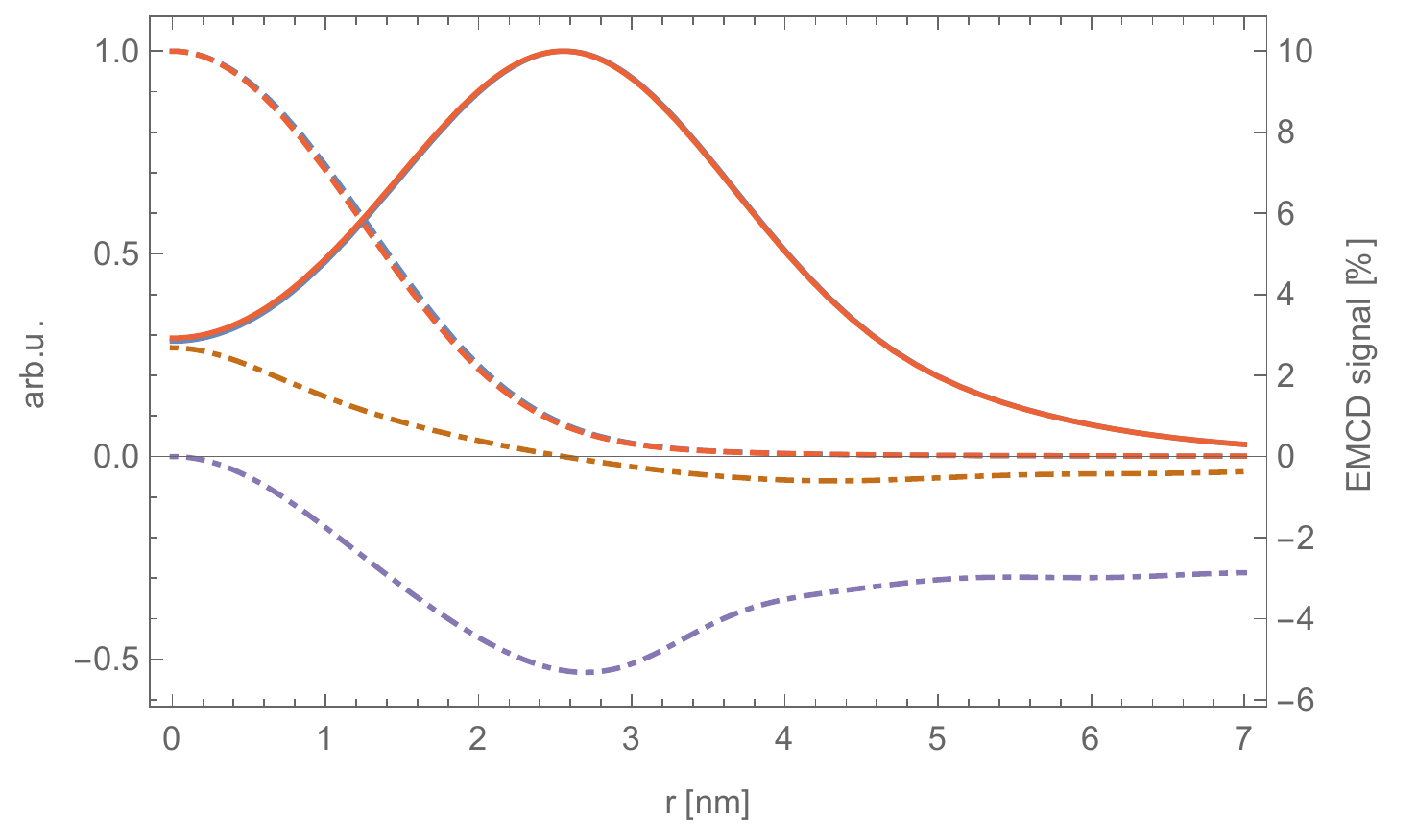} 
	\caption{Incoherent broadening: Radial intensity profiles of $m=\pm 1$ filtered images as in Fig.~\ref{fig:CoCoherent} with a defocus of \unit{0}{\micro\meter} (dashed lines) and \unit{4}{\micro\meter} (solid lines), respectively, for an illuminated specimen area of \unit{\sim1.9}{\nano\meter} ($\sigma = 0.8$) and a $C_s$ of \unit{1.2}{\milli\meter}, according to Eq.~\ref{ImBroadened}. Due to the incoherent broadening effect the focused vortices no longer show the characteristic dip in the center and the defocused radial profiles are hardly distinguishable. The resulting EMCD signal (dot-dashed curves) is dramatically lowered and amounts to $\sim 3\%$ in the central region of the defocused vortices and $\sim 5\%$ approximately at the beam waist for the focused ones. (Note the change of the right scale.)}
	\label{fig:CoIncoherent}
\end{figure}

The observed vortices are calculated with Eq.~\ref{FT3}, but now including the defocus $df$ and the spherical aberration $C_s$:
\begin{multline}
 \psi_{m \mu}({\bf r})= 
 \frac{i^{m+\mu}}{2 \pi}\, e^{-i(m+\mu)\varphi_r}\int_0^{q_{mask}} \tilde \psi_\mu(q)\, J_{|m+\mu|}(q\, r)\, \\
 e^{i({df q^2}/{2 k_0}+{C_s q^4}/{4 k_0^3})}\,  q \,dq .
\label{FT4}
\end{multline}
When a homogeneous specimen is illuminated all atoms within the beam will contribute incoherently with their respective signals. This incoherent broadening effect according to the finite illuminated area of the specimen is taken into account by a convolution with a Gaussian as described in~\cite{SchattUm2012a}. Thus, the final simulated radial intensity distribution is given by
\begin{equation}
 I_{m}^{\sigma}({r})= 
 e^{-(1/2)(r/\sigma)^2}\int_0^{\infty} \psi_{m \mu}(r')\, e^{-(1/2)(r'/\sigma)^2} I_{0}\left(\frac{r r'}{\sigma^2}\right)\,  r' \,dr', 
\label{ImBroadened}
\end{equation}
where $I_0$ represents the modified Bessel function of first kind of order zero and $\sigma$ the amount of incoherent broadening. The resulting illuminated area (FWHM) at the specimen is $\sim2.4\, \sigma$. This incoherent broadening effect will reduce the EMCD signal as shown in Fig.~\ref{fig:CoIncoherent}. 
But still, the defocused case is preferable because the tiny differences in width for the focused case are hardly observable, see Fig.~\ref{fig:CoIncoherent}.
In Eq.~\ref{ImBroadened}, there are two free parameters, defocus and broadening width, which are used to obtain best fits to the experimental data shown in the next section.

\section{Experimental}
\label{sec:Experimental}

The fundamental method and scattering geometry elaborated above have been realized in a proof-of-principle experiment on a FEI TECNAI F20 TEM equipped with a GATAN GIF Tridiem spectrometer (GIF) and a high-brightness XFEG. The acceleration voltage was set to \unit{200}{\kilo\volt}, whereas the condenser system was set up in a way to achieve a high beam current at a sufficiently small spot size, i.e. providing a beam current of \unit{\sim500}{\pico\ampere} incident on the sample in a \unit{\sim 1.9}{\nano\meter} probe (FWHM) with a convergence semi-angle of \unit{3.8}{\milli\radian}.

\begin{figure}[ht]
	\centering
	\includegraphics[width=45mm]{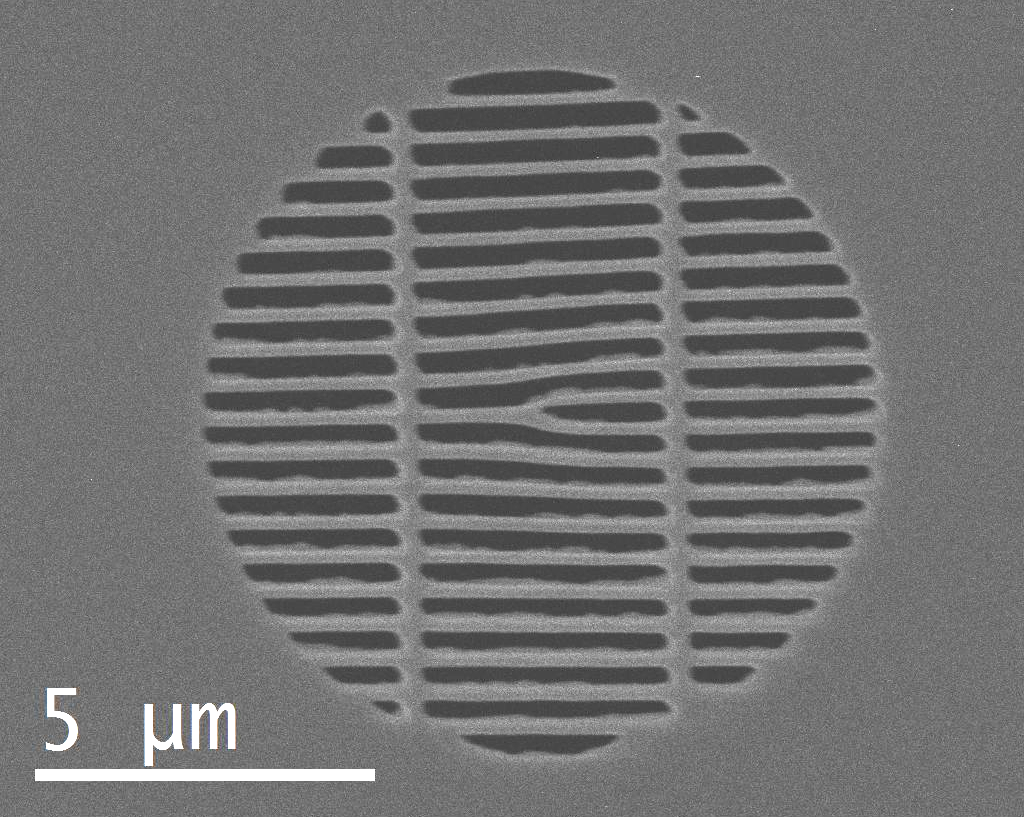} 
	\caption{Holographic fork mask prepared by FIB milling used as a vortex filter in the SAA holder. The mask has a diameter of \unit{10}{\micro\meter} with a grating periodicity of \unit{500}{\nano\meter}, giving a diffraction of the first vortex orders at $2\theta_B=\unit{5}{\micro\radian}$.}
	\label{fig:VortexMask}
\end{figure} 

\begin{figure}[ht]
	\centering
	\includegraphics[width=90mm]{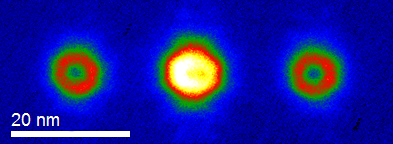} 
	\caption{Experimental image showing the vortex orders $m=+1$ (left), $m=-1$ (right) and $m=0$ (middle), produced by the SAA vortex filter at the Co $L_3$-edge defocused by \unit{4}{\micro\meter}. 
}
	\label{fig:CoExpColor}
\end{figure}

Fig.~\ref{fig:VortexMask} shows the vortex filtering holographic fork mask that was placed in the SAA holder. It was produced by FIB milling into a \unit{300}{\nano\meter} Pt layer deposited on a \unit{200}{\nano\meter} $\text{Si}_3 \text{N}_4$ membrane.   
With a diameter of \unit{10}{\micro\meter} and a grating periodicity of \unit{500}{\nano\meter} (back-projected: \unit{9.4}{\nano\meter}), it exhibits a Bragg angle of $\theta_B=\unit{5}{\micro\radian}$ separating the central spot from the first vortex orders in the eucentric plane by $\sim \unit{20}{\nano\meter}$\footnote{The separation distance was calculated using $2\theta_{Bragg}\, dz$, with a camera-length of $dz=\unit{75}{\micro\meter}$ and the back-projected grating periodicity $g=\unit{9.4}{\nano\meter}$.}. As a result the vortex orders $\pm1$ are still well separated from the central peak for defocus values of $df = \unit{4}{\micro\meter}$ and higher, see Fig.~\ref{fig:CoExpColor}.

For the preparation of the Co sample, a 70 nm thin Co layer is deposited onto a NaCl crystal. The thin Co foil is then extracted by dissolving the NaCl in water. Afterwards, the Co foil is netted with a commercially available Cu grid, resulting in a free standing nano-crystalline Co film of \unit{70}{\nano\meter} thickness,  with randomly oriented \unit{20}{\nano\meter} crystallites.
In the following section, the experimental setup of the vortex filter experiment on the Co film is described in detail.

In imaging mode, the objective lens is set to the eucentric focus value with the sample in the eucentric height. From that position, the Co sample is then lifted by $dz=\unit{75}{\micro\meter}$. The beam is focused onto the lifted specimen using C2 excitation and observing the ronchigram in the eucentric plane.
The microscope is set to the diffraction mode with a camera length of \unit{39.5}{\meter} (including the GIF magnification), which is necessary to resolve the \unit{5}{\micro\radian} separated EVBs. Focusing is solely done with the diffraction lens. Then, all three vortex orders are imaged at the GIF camera in the energy filtered TEM mode\footnote{In fact we are working in the energy filtered selected area diffraction (EFSAD) mode~\cite{LoefflerEnnenTianEtAl2011} because the microscope is set to diffraction mode and the SAA (vortex filter mask) is used.}, operated at the edge threshold energy of the $L_3$-edge of Co of \unit{780}{\electronvolt} using a slit width of \unit{15}{\electronvolt}, see Fig.~\ref{fig:CoExpColor}.
In order to keep the illumination conditions constant, the drift tube was used to adjust the spectrometer to the desired Co $L_3$-edge energy of \unit{780}{\electronvolt} instead of the high tension.

\begin{figure}[ht]
	\centering
	\includegraphics[width=90mm]{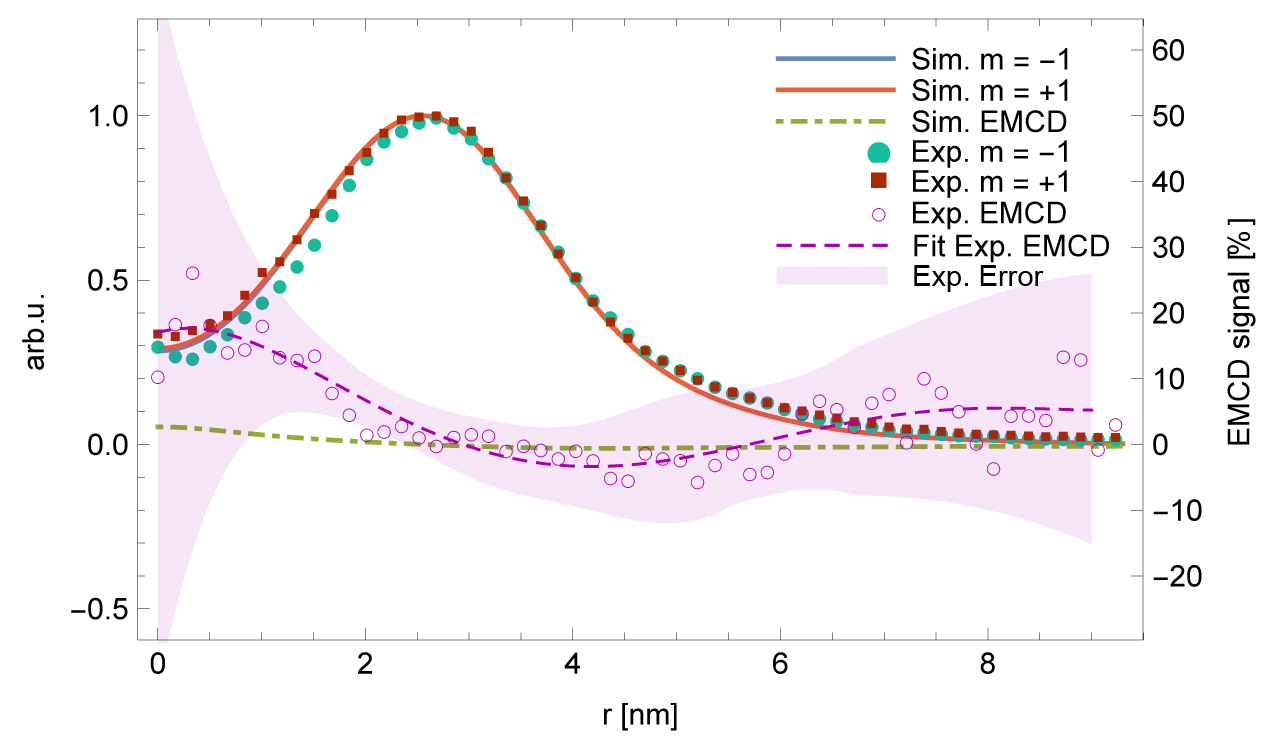} 
	\caption{Rotational averages of the outermost vortex orders in Fig.~\ref{fig:CoExpVsSimulation} (red and green full dots) and their best fits using Eq.~\ref{ImBroadened} including incoherent broadening with $\sigma=\unit{0.8}{\nano\meter}$ and a defocus of $df=\unit{4}{\micro\meter}$ (full lines). The theoretically predicted EMCD signal (green, dot-dashed line) is compared to the experimental one (magenta open circles and a dashed polynomial fit curve). The error in the EMCD signal, given by the magenta shaded area, representing the RMS value including Gaussian error propagation, indicates that the faint EMCD signal cannot be discerned under present experimental conditions.}
	\label{fig:CoExpVsSimulation}
\end{figure}

Fig. 8 shows the experimental energy filtered image of the vortex aperture from electrons which have transferred \unit{780}{\electronvolt} to the Co sample. Due to the extremely low count rates, Fig.~\ref{fig:CoExpColor} is acquired taking four frames with an acquisition time of 100 s per frame with four fold binning. Subsequently, the frames are stacked and aligned using Image J~\cite{SchneiderRasbandEliceiri2012}. To extract the radial intensity profiles given in Fig.~\ref{fig:CoExpVsSimulation}, {\it Digital Micrograph} scripts are used to determine the exact vortex orders' centres, cropping them and doing the rotational (azimuthal) average. 

Fig.~\ref{fig:CoExpVsSimulation} shows that the simulation\footnote{For the sake of simplicity, elastic scattering inside the nano-crystalline structure of the specimen was not taken into account in the single atom approach given here.} with the chosen parameters is in very good agreement with the experimental data. Curiously, the experimental radial profiles show a strong difference in the central region ($\sim15\%$) which is similar to classical EMCD measurements~\cite{Schattschneider2012} and previous vortex filter EMCD experiments~\cite{VerbeeckNature2010}. However, the simulation predicts a much smaller EMCD signal ($\sim3\%$). This discrepancy is possibly due to (i) skew optic axes  which gives rise to slight differences in apparent defocus $df$ for the positive and negative vortex orders, (ii) artefacts from the mask production~\cite{BarwickGronnigerYuanEtAl2006} and (iii) OAM impurities stemming from the SAA vortex mask~\cite{ClarkBecheGuzzinatiEtAl2014}.

Fig.~\ref{fig:CoExpVsSimulation} also clearly shows that the EMCD signal is much smaller than the relative root-mean-square (RMS) value of the experimental EMCD signal ($\sim\pm40\%$, magenta shaded region) and thus cannot be detected reliably under present experimental conditions. 
Since the profiles are azimuthally averaged, the absolute signal-to-noise ratio (SNR) per radial position is best for the largest radius where an average over 512 pixels was taken, and lowest for the second point (9 pixel average). This was taken into account numerically for the results shown in Fig.~\ref{fig:CoExpVsSimulation}. The magenta shaded region in Fig.~\ref{fig:CoExpVsSimulation} is calculated using Gaussian error propagation, thus showing the relative RMS value in the azimuthal direction of the EMCD signal defined by Eq.~\ref{EMCD}.

Furthermore, the error in Fig.~\ref{fig:CoExpVsSimulation} (magenta shaded region) only represents the statistical error; systematic errors such as beam-drift, beam damage, artefacts due to non-isotropic vortex rings etc. have not been included.
In practice, the sample's spin polarization may be less than 100 \%, and the vorticity may change during the propagation of the outgoing vortex beams through the specimen~\cite{LoefflerSchattschneider2012}, thereby decreasing the expected EMCD signal as well. 

In view of these results and considerations, further investigations and improvement of the experimental conditions are necessary to proof the applicability and reliability of the EMCD vortex filter method.

\section{Conclusions}
In this work we investigated the feasibility of detecting an EMCD signal when incorporating a fork mask as a vortex filter in the SAA plane in a standard two-condenser lens field emission TEM. By lifting the sample far above the eucentric position, a vortex filter mask in the SAA plane can be properly illuminated. Thus, it produces well separated vortex orders which should, in principle, carry the EMCD information in the asymmetry of their respective central intensities. This method could become a promising method for studying magnetic properties of amorphous or nanocrystalline materials, which is impossible in the classical EMCD setup. 
So far the experimental tests show that the SNR is still too low and that for a successful experimental realization substantial progress in the experimental conditions is compulsory. For example, to improve the SNR future experiments should incorporate larger SAA fork masks, e.g. at least $30$ to \unit{50}{\micro\meter} in diameter. As the collected signal scales with the mask area, the acquisition times could then be lowered by an order of magnitude. Also, incorporating a HM in the contrast aperture holder located in the DP would simplify the experimental setup as well as increase the collection efficiency. 
Finally, using state-of-the-art aberration corrected microscopes it is possible to increase the lateral coherence of the probe beam while at the same time keeping the beam current high. This would enhance the EMCD signal strength by an order of magnitude. Thus, in the light of above considerations and the proof-of-principle experiment, detection of EMCD signals using HM as a vorticity filter seems to be feasible but needs thorough control of experimental parameters like spot size, vortex masks fidelity, sample- and system stability.

\section{Acknowledgements}
T.S. acknowledges financial support by the Austrian Science Fund (FWF), project I543-N20 and by the European research council, project ERC-StG-306447, P.S. acknowledges financial support by the FWF, project I543-N20 and S.L. acknowledges financial support by the FWF, project J3732-N27.

\bibliography{Lit2016TS,EMCD,NewRefs2,mybibfile,bibInga1,scopus-4,papers}

\end{document}